\documentclass{iau} 
\usepackage{graphicx}

\title[Uncertain Photometric Redshifts with Deep Learning Methods] %% give here short title %%
{Uncertain Photometric Redshifts with Deep Learning Methods}

\author[A. D'Isanto]   %% give here short author list %%
{A. D'Isanto$^1$
 \thanks{The author gratefully acknowledges the support of the Klaus Tschira Foundation.}}

\affiliation{$^1$Heidelberg Institute for Theoretical Studies (HITS)\\
Schloss-Wolfsbrunnenweg 35, 69118 Heidelberg - GERMANY \\ email: {\tt antonio.disanto@h-its.org} \\[\affilskip]
}

\pubyear{2016}
\volume{325}  %% insert here IAU Symposium No.
\jname{Astroinformatics 2016}
\editors{A.C. Editor, B.D. Editor \& C.E. Editor, eds.}
\begin{document}

\maketitle

\begin{abstract}
The need for accurate photometric redshifts estimation is a topic that has fundamental importance in Astronomy, due to the necessity of efficiently obtaining redshift information without the need of spectroscopic analysis. We propose a method for determining accurate multi-modal photo-z probability density functions (PDFs) using Mixture Density Networks (MDN) and Deep Convolutional Networks (DCN). A comparison with a Random Forest (RF) is performed. 
\end{abstract}

\firstsection % if your document starts with a section,
\section{Introduction}
Determination of distances for astronomical objects through redshift acquired in the recent years an increasingly importance, having a fundamental role in cosmological research.
In fact, it is well known that redshift is a fundamental step of the cosmic distance ladder.
Redshift is traditionally obtained through spectroscopic analysis but due to long integration times and costly instrumentation requirements, it is not possible to measure it for all objects.
Therefore, a convenient alternative is the estimation of photometric redshifts, e.g. based on measurements of pure photometry.
However, the uncertainty of such an approach is much higher than the measurement errors obtained from spectroscopy.
For this reason, the astronomical community has focused in the uncertainty quantification of redshift estimates through probability density functions (PDFs), instead of using simple point estimates.
In this work we propose two neural network models based on Mixture Density Networks (MDN) (\cite[Bishop 1994]{bishop}).
We use a deep MDN as first architecture, designed to use photometric features as inputs and to generate PDFs. 
The second architecture is a combination of a Deep Convolutional Network (DCN) (\cite[LeCun et al. 1998]{lecun}) with a MDN with the purpose to obtain photo-z PDFs based on images as input.
We will show that this approach achieves better predictions due to its use of image data that - in contrast to using pre-defined features - allows to capture more details of the objects.
We compare the results obtained with a commonly used tool in the related literature, the Random Forest (RF) (\cite[Breiman 2001]{breiman}).

\section{Deep learning algorithms}\label{sec:algorithms}
In the next two subsections we give a description of the deep learning algorithms used for the experiments.
\subsection{Mixture Density Network}\label{sec:mdn}
A Mixture Density Network (\cite[Bishop 1994]{bishop}) is the combination of a feed-forward neural network and a Gaussian mixture model. The outputs of the network parametrize the Gaussian mixture $p(\theta|x) = \sum_{j=1}^{n}\omega_{j}\mathcal{N}(\mu_{j}, \sigma_{j})$, i.e. they define the means, variances, and weights.
Thus the MDN produces a multi-modal PDF suitable for the case of photo-z, which a flexible enough to represent a multi-modal behavior.
The means, variances and weights, are then obtained by the outputs $z$ of the network:

\begin{equation}\label{eq:mu_sigma_omega}
\mu_{j} = z_{j}^{\mu}\ ,
\hspace{1cm}
\sigma_{j} = \exp(z_{j}^{\sigma})\ ,
\hspace{1cm}
\omega_{j} = \frac{\exp(z_{j}^{\omega})}{\sum_{i=1}^{n}\exp(z_{i}^{\omega})}\ .
\end{equation}

\noindent Normally the MDN uses negative log-likelihood as a loss function, but in this work we use the \emph{continuous rank probability score} (CRPS) (\cite[Gneiting et al. 2005]{gneiting}) as loss function.
This is to obtain a trained MDN which produces PDFs both well calibrated and sharp as measured by the CRPS, as explained in detail in \cite[Polsterer et al. (2016)]{polsterer}.

\subsection{Deep Convolutional Network}\label{sec:dcmdn}
A Deep Convolutional Network is a model in which several convolutional and sub-sampling layers are coupled with a fully-connected network.
This architecture is particularly meant to learn from raw image data.
In our case, we want to estimate redshifts directly from images, without the need to extract photometric features, so we couple a DCN with a MDN, in order to produce photo-z PDFs directly from SDSS images.
We alternate convolutional and pooling layers to generate feature maps and generate a hierarchically compressed representation of the input data.
The output of the convolutional network is then taken as input for the MDN which produces a multi-modal predictive density for photo-z.
Thereby the extraction of the \emph{feature maps} is automatically done by the network.
Those obtained feature maps are then taken as inputs for the fully-connected part.
We choose a modified version of the LeNet-5 architecture (\cite[LeCun et al. 1998]{lecun}), properly coupled with the presented MDN (see Section~\ref{sec:mdn}), obtaining what we call a Deep Convolutional Mixture Density Network (DCMDN).
In Tab.~\ref{tab:architectures} there is the architecture of the DCMDN used for the experiments, designed to run on GPUs, using a cluster equipped with Nvidia Titan X.

\begin{table}\footnotesize
\centering
\begin{tabular}{| c | c | c | c | c |}
\hline
\# & Type & Size & Maps & Activ \\
\hline
1 & input & 28x28 & / & / \\
\hline 
2 & Conv & 3x3 & 256 & tanh \\
\hline
3 & Pool & 2x2 & 256 & tanh \\
\hline
4 & Conv & 2x2 & 512 & tanh \\
\hline
5 & Pool & 2x2 & 512 & tanh \\
\hline
6 & Conv & 3x3 & 512 & ReLu \\
\hline
7 & Conv & 2x2 & 1024 & ReLu \\
\hline
8 & MDN & 500 & / &  tanh \\
\hline
9 & MDN & 100 & / & tanh \\
\hline
10 & output & 15 & / & Eq. \ref{eq:mu_sigma_omega} \\
\hline
\end{tabular}
\caption{DCMDN architecture}
\label{tab:architectures}
\end{table}

\section{Experiments and Analysis}\label{sec:experiments}
The data we use for the experiments are taken from the Sloan Digital Sky Survey Quasar Catalog V (\cite[Richards et al. 2010]{richards}), based on the 7-th data release of the Sloan Digital Sky Survey (SDSS), consisting in $105,783$ spectroscopically confirmed quasars, in a redshift range between $0.065$ and $5.46$.  
For the experiments we use a random subsample of $50,000$ patterns.
For each pattern we take the five \emph{ugriz} magnitudes as input features and the respective images in the same bands.
Finally, we compare the performances of MDN and DCMDN with the widely used RF. 

The RF, in its original architecture, is not meant to produce PDFs.
In order to obtain a distribution, we first collect the predictions $z_{t,n}$ of each individual decision tree $t$ in the forest, for every $n$-th data item.
We take $T=256$ number of trees in the forest and define the PDF for the RF by fitting a mixture of 5 Gaussian components to the outputs, $p(\theta|x)=\sum_{j=1}^{5}\omega_{j}\mathcal{N}(\theta|(\mu_{j},\sigma_{j}))$, as we described also in Section~\ref{sec:mdn} for the MDN.

\begin{figure*}[h!]
\centering
 \includegraphics[width=\textwidth]{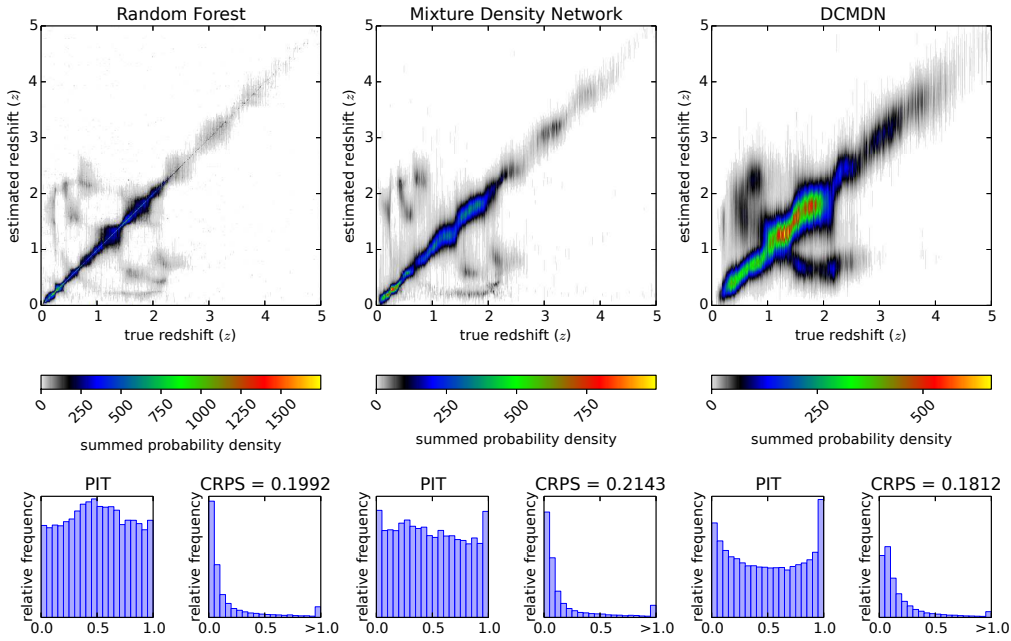}
    \caption{Results of the predictions obtained with the MDN and the DCMDN, compared with the RF results.
    For each experiment, three plots are given.
The upper plots compare the spectroscopic redshift with the predictive distributions produced by the models, where the color indicates the summed probability density of the distributions.
In the two lower plots, the histogram of the PIT values and the histogram of the individual CRPS values, are shown.
The mean CRPS value is also given.
    }
    \label{fig:results}
\end{figure*} 

For the RF and the MDN we use as input the 5 magnitudes plus all the possible color combinations, obtaining a 15-dimensional feature vector, respectively.
The generated training and test set both contain $25,000$ patterns.
The DCMDN is trained on the images, that are obtained using the \emph{Hierarchical Progressive Surveys} data partitioning format (\cite[Fernique et al. 2015]{fernique}) and performing a proper cutout on client side, in order to obtain the desired dimensions (28x28).
Each pattern is originally a stack of 5 images in the \emph{ugriz} filters, where every pixel is converted from flux units to \emph{luptitudes} (\cite[Lupton et al. 1999]{lupton}).
As done with the usual features, we additionally form the color images from the \emph{ugriz} images by taking all possible pairwise differences, thus obtaining a stack of 15 images; every object/pattern is then represented by a tensor of dimensions 15x28x28.
In order to have a rotational invariant network, we perform data augmentation, taking rotations of each image at 0, 90, 180, 270 degrees.
By doing so, we obtain a training set of $100,000$ images, a validation set of $50,000$ images and a test set of $50,000$ images.
Dropout is also applied to limit overfitting.

The results of the experiments are reported in Fig.~\ref{fig:results}.
Following \cite[Polsterer et al. (2016)]{polsterer}, we use two statistical tools: the CRPS as a score function, and the \emph{probability integral transform} (PIT) histogram (\cite[Gneiting et al. 2005]{gneiting}), in order to obtain a visual estimation of the quality of the produced PDFs.
In the RF experiment, the model reaches a CRPS of 0.20 and the PIT shows a bit of overdispersion.
The performance of the MDN is a bit worse than the RF in terms of the CRPS, with a score of 0.21, but exhibits a better calibrated PIT.
Using the DCMDN architecture we achieve the best results in terms of the CRPS, with a score of 0.19.
The resulting PIT is acceptable, although it is still showing some underdispersion.
The reason for the better overall performance of the DCMDN is that the features-based approach use only a fraction of the available information.
In fact, in the process of features extraction a lot of information gets lost.
Instead, using images, the DCMDN is able to automatically determine thousands of features, leading to a better prediction of the photo-z PDFs.

\section{Conclusions}\label{sec:conclusions}
Main purpose of this work is to show a method to produce photo-z PDFs using deep learning architectures.
We generate very good probabilistic predictions based on features or images as input, producing a Gaussian mixture model as output. 
Our proposed architectures show better performances in the comparison carried out with a RF based method.  
In particular, we show that the proposed DCMDN gives the best performance, as it is able to use the entire information contained in the images.
As showed by the PIT analysis, some optimization with respect to calibration can still be done, in order to deal with some dispersion phenomena.
We firmly believe that the presented method needs little improvements to become a standard in predicting photo-z PDFs.
As regression problems are very common in Astronomy, this approach can easily be applied to many other scientific topics.

%\begin{discussion}

%\end{discussion}

\end{document}